\pdfoutput=1

\documentclass[useAMS,usenatbib]{mn2e}
\title[Gravity estimate of Column Density]{Reliable estimation of the column
density in Smoothed Particle Hydrodynamic simulations.}
\author[Young et al.]{
M.D.~Young$^1$, 
E.~Bertram$^{1,2}$, 
N.~Moeckel$^1$ and
C.J. Clarke$^1$ \\
$^1$Institute of Astronomy, University of Cambridge, Madingley Road, Cambridge,
CB3 0HA, United Kingdom\\
$^2$Zentrum f\"ur Astronomie der Universit\"at Heidelberg, Institut f\"ur
Theoretische Astrophysik, Albert-Ueberle-Str. 2, 69120 Heidelberg,
Germany}
\usepackage[pdftex]{graphicx}
\usepackage{natbib}
\usepackage{url}

\begin{document}
\date{Written June 12th 2012}
\maketitle
\begin{abstract}
We describe a simple method for estimating the vertical column density in
Smoothed Particle Hydrodynamics (SPH) simulations of discs.  As in the method
of \cite{PolytropicCooling}, the column density is estimated using
pre-computed local quantities and is then used to estimate the radiative
cooling rate.  The cooling rate is a quantity of considerable importance, for
example, in assessing the probability of disc fragmentation.  Our method
has three steps: (i) the column density from the particle to the mid plane is
estimated using the vertical component of the gravitational acceleration, (ii) 
the ``total surface density'' from the mid plane to the surface of the disc is
calculated,
(iii) the column density from each particle to the surface is calculated from
the difference between (i)
and (ii). This method is shown to
greatly improve the accuracy of column density estimates in disc geometry
compared with the method of Stamatellos.  On the other hand, although the
accuracy of our method is still acceptable in the case of high density
fragments formed within discs, we find that the Stamatellos method performs
better than our method in this regime.  Thus, a hybrid method (where the
method is switched in regions of large over-density) may be optimal.
\end{abstract}

\section{Introduction}

Smooth Particle Hydrodynamics (SPH) \citep{LucySPH,MonaghanSPH} is a Lagrangian technique for
simulating fluid flows using a particle representation.  This technique assigns
each particle a mass, position and internal energy and then interpolates state
variables, such as density, by ``smoothing'' over neighbouring particles.
Gravity has been incorporated into the SPH formalism, allowing SPH to be
used to simulate astrophysical fluids.  However, the thermal evolution of many
astrophysical systems is often governed by radiative transfer effects, in
addition to hydrodynamic energy transfer.  As an accurate description of a
system's thermal evolution is of vital importance in many astrophysical systems
(e.g. accretion discs, stellar gas clouds) a lot of effort has recently been
made to add radiative transfer to SPH
\citep{RadiativeSPH1,RadiativeSPH2,PolytropicCooling,HybridCooling}.  Unfortunately, a full three
dimensional, frequency dependent description of radiative transfer is currently
computationally impossible, so simplifying assumptions have to be made.

Since SPH is a Lagrangian method, the purpose of modeling radiative transfer
effects is to provide a net cooling (or heating) rate {\it per particle} as a
result of radiative processes.  For example, a popular choice for
simulating self-gravitating discs is to use a cooling time prescription, where
the radiative cooling rate $\dot{U}$ is set equal to $\frac{-U}{t_{cool}}$ where
$U$ is the cooling rate per unit mass and $t_{cool}$ is simply 
parameterised as a prescribed 
multiple of the local dynamical timescale \citep{BetaCooling}.  Although such a description grossly
oversimplifies the underlying physics, it has a very low computational cost and
so simulations including approximating radiative cooling can be run without sacrificing
spatial or
temporal resolution.  Recently, Stamatellos et al have proposed a method that
improves upon the cooling time prescription without significantly increasing
the computational cost \citep{PolytropicCooling}.  Forgan et al extended this method to include
heat transfer between particles, by combining the Stamatellos method with the
flux limited diffusion (FLD) method \citep{HybridCooling}.

The Stamatellos method estimates the optical depth by assuming that the
relationship between the two locally computed variables, density and
gravitational potential, and the optical depth is the same as it is for a
mass-weighted average of that relationship over a self-gravitating polytropic
sphere. They then estimate the 
local cooling rate using only the local temperature and optical depth.  This
estimate tends to the radiative diffusion approximation at high optical depths
but does not involve calculating noisy derivatives as is required in a proper
implementation of radiative diffusion \citep{SPHFLD}.
The aim of such methods is not to achieve perfect agreement with the
full radiative transfer equations, but to achieve an accuracy that is
comparable in magnitude to the other uncertainties such as those associated
with the grey approximation and the appropriate values of the frequency
averaged opacity.

However, although the Stamatellos method has proved successful for modelling spherical
systems, Wilkins \& Clarke have shown that it can systematically underestimate 
the cooling in disc geometries by as much as a factor of four in the region of the 
mid plane, where column density estimation is critical to estimating the cooling
\citep{WilkinsAndClarke}.  This underestimate stems from the Stamatellos method
overestimating the column density, $\Sigma$, which appears as a
quadratic term in the expression for the cooling rate in the optically thick
limit.  An
example of where radiative transfer is important to the evolution of a disc
system is the study of gravitational instabilities in proto-planetary disc.
The cooling time prescription has been used to study these discs in great
detail, but a more accurate description of the cooling in such systems is
necessary to improve our understanding of these gravitational instabilities
\citep{DataPaper}.

In this paper we propose a variation on the Stamatellos method for
calculating the cooling rate for the special case of disc geometries, by
improving the estimate of the column density in this case.
Our method still only requires the use of local quantities, already calculated
by the gravitational and hydrodynamic codes and as such remains computationally
inexpensive. The paper is organized as follows.
In section \ref{sec:method}, we describe our new method in detail. In section \ref{sec:analtests},
we test our method on a series of discs for which all quantities of interest 
can be obtained analytically or semi-analytically.
In section \ref{sec:realdiscs} we test our method on realistic disc
simulations
that have been evolved long enough to either fragment or reach marginal stability.  
Finally, section \ref{sec:conclusion} summarises our conclusions.

\section{The Method}
\label{sec:method}

For a typical astrophysical disc, most of the cooling occurs at the disc's
surface.  Therefore, we assume that the greatest contribution to the radiative
cooling of a particle within such a disc comes from energy radiated vertically
out of the disc, rather than along the optically thick mid-plane.
Assuming this is true, we estimate the cooling rate of each particle using the
optical depth,
$\tau$, along a vertical path from the particle to the disc's surface. The true optical
depth is given by $\tau=\int \kappa (z) \rho (z) dz$. In order to avoid
evaluating this costly integral we make the approximation that $\tau =\kappa
\int \rho dz = \kappa \Sigma$, where $\Sigma$ is the column density to the
surface and $\kappa$ is the \emph{local} opacity. Next, we follow
Stamatellos et al in defining the cooling rate per unit mass to be:

\begin{equation}
  \dot{U} = \frac{\sigma (T^4 - T_0^4)}{\tau \Sigma +
  \widetilde{\kappa}^{-1}}
  \label{eq:cooling}
\end{equation}

\noindent where, $\tau$ is the optical depth to the surface of the disc, $\sigma$ is the
Stefan Boltzman constant, $\Sigma$ is the column density from the particle to the surface of the
disc and $\widetilde{\kappa}$ is the local Planck-mean opacity.  The $T_0$ term is a background
temperature below which particles are not allowed to cool.  
%This temperature
%can used to include heating from a central star.  For example, Stamatellos et al. use:

%\begin{equation}
%  T_0^4 = (10 K)^4 + \Sigma_* \left\{ \frac{L_*}{16 \pi \sigma |r-r_*|} \right\}
%  \label{eq:starcool}
%\end{equation}

%where $L_*$ and $r_*$ and the stars luminosity and position.  

In the regime where $T^4 \gg T_0^4$, we find that there are two limiting cases.
The first limiting case is the optically thin limit, where $\tau\Sigma \ll
\widetilde{\kappa}^{-1}$.  In this limit, the cooling simply reduces to

\begin{equation}
  \dot{U} =  \sigma T^4 \widetilde{\kappa}
  \label{eq:opticallythin}
\end{equation}

\noindent which is just the cooling rate for an isolated particle, in an environment with
opacity $\widetilde{\kappa}$.  However, note that this is \emph{not}
a good approximation to the cooling rate per unit mass for an optically
thin layer on top of an optically thick disc, which is a commonly
encountered  situation. Fortunately, in practice the thermodynamics of such
layers are often controlled by the background temperature, $T_0$. 

In the optically thick limit, $\tau\Sigma \gg \widetilde{\kappa}^{-1}$ and
equation \ref{eq:cooling} becomes

\begin{equation}
  \dot{U} = \frac{ \sigma T^4}{\Sigma \tau}
  \label{eq:opticallythick}
\end{equation}

\noindent which is an approximation to the commonly used diffusion approximation (see
\cite{MIHALAS}, section 2.3). To see why, consider the optically thick limit, in which
the radiative flux is given by:

\begin{equation}
F=-\frac{4}{3\kappa\rho} \nabla\sigma T^4 
\end{equation}

\noindent and the corresponding cooling rate per unit mass as:

\begin{equation}
  \dot{U} = \frac{1}{\rho}\nabla . F 
\end{equation}

If integration and differentiation over $z$ are replaced by simple
multiplication and division by an effective vertical scale height, $H$,
then an expression of the same form as \ref{eq:opticallythick} is obtained.

\begin{equation}
  \dot{U} = \frac{1}{\rho} \nabla . F \approx \frac{\sigma T^4}{\kappa \Sigma^2}
  \approx \frac{\sigma T^4}{\tau \Sigma}
  \label{eq:limitCase}
\end{equation}

Note that because $\dot{U} \propto \Sigma^{-2}$, any inaccuracy in
the column density to the surface, $\Sigma$, will have a large effect on 
the inaccuracy of the cooling rate. 

$\Sigma$ is calculated in a three step
process.  Firstly, an estimate of the column density between the particle and
the mid-plane is obtained. This is done using only the vertical component of
the gravitational
acceleration, which is already calculated by the simulation making this step
essentially ``free'' from a computational standpoint.  Secondly, a total surface
density map (i.e. map of column density from the mid-plane to the surface)
is computed. Finally, the column
density to the surface is calculated by subtracting the column density to the
mid-plane from the total surface density.

\subsection{Estimating the column density to the mid-plane}

We estimate  the column density of each particle to the mid-plane using the
approximation: 

\begin{equation}
  g_z = -\frac{GMz}{r^{3}} - 4\pi G \Sigma ' sign(z)
  \label{gravity}
\end{equation}

\noindent where $M$ is the mass of the central star, $r=\sqrt{x^2+y^2+z^2}$ and $\Sigma'$
is the column density \emph{between the point at height $z$ and the mid-plane}.
The first term in this expression is the
contribution from the central star, whereas the second term comes from assuming
that the disc is infinite in extent and has the same vertical density structure
everywhere.  In reality, there will be further corrections to this expression
due to radial variation in the disc's vertical density structure
and its finite extent (see e.g. Appendix of \cite{AnalyticDiscGravity}.)  
However, we expect these contributions to be
small in most cases, an assumption that will be
investigated in detail in section \ref{sec:analtests}.

As the gravitational acceleration, $g_z$, is already calculated by the simulation
for every particle, we can re-arrange equation (\ref{gravity}) to estimate the
column density from the particle to the mid-plane\footnote{Note that this
assumes that the value of $g_z$ calculated by the code includes the
contribution from the star.  If it does not, the term due to the star's
potential should be dropped from equations \ref{gravity} and \ref{sigmap}.}.

\begin{equation}
  \Sigma' = -\frac{sign(z)}{4\pi G}\left(g_z + \frac{GMz}{r^3}  \right)
  \label{sigmap}
\end{equation}

\subsection{Constructing a surface density map}

In order to calculate the cooling, we need to know the column density
between the particle and the surface of the disc, and therefore the
column density to the mid-plane, estimated in equation (\ref{sigmap}) above,
has to be subtracted from an estimate of the total column density.  
This map can be calculated in a number of different ways,
discussed at length in Appendix \ref{sec:surfaceDen} and the section below. 
In this paper, we calculate the map by projecting all particles
onto the mid-plane and then interpolating for each particle from
a density map evaluated at the location of a subset of particles ($10 \%$).  The 
surface density for the remaining 90\% of points are interpolated when needed.  
Once obtained, we can estimate the column density to the surface, $\Sigma$, by:

\begin{equation}
  \Sigma = \Sigma_{map}-\Sigma'
  \label{totalCD}
\end{equation}

\subsection{Computational efficiency}
\label{sec:compeff}

The construction of a total surface density map,
has the potential to be computationally expensive.  The potentially expensive
aspect of this step is the estimation of the surface density at each particle
from the particle positions and masses, projected onto the mid plane.
There already exists an extensive literature devoted to solving this problem,
which we will not attempt to reproduce in detail here. Importantly, the choice 
of method for estimating density involves
a trade off between accuracy, computational speed and the ability to recover
sharp density gradients (see \citep{Ferdosi} for a discussion of these trade-offs
in the context of astronomical datasets).  

SPH density estimation methods give excellent accuracy, but a high computational cost
($O(N^2)$, e.g. the ``DEDICA'' method
in \cite{Ferdosi})).  At the other extreme, grid based methods 
offer excellent computational efficiency ($O(N)$), but with
reduced accuracy (e.g. the ``MBE'' method in
\cite{Ferdosi}). The method we use here (describe in appendix \ref{sec:surfaceDen}) is
intermediate in complexity between these two extremes and is very similar to the
``kNN'' method described in \cite{Ferdosi}, which has $O(NlogN)$ scaling.

The computational cost of any of the above method can be
reduced, at the cost of ``smoothing out'' the density distribution, 
by evaluating the density at a subset of points and then interpolating
the values at the particle locations.

It is important to note that the calculation of a total surface density map is 
a two-dimensional version of the calculation of the three-dimensional density of each particle,
which must be performed by any SPH code.  As such, even in the worst case
scenario this step will be no more expensive than the 3D density calculation
already performed by the code.  Furthermore, as there is little gain in calculating the surface
density to greater accuracy than the other sources of error in our method
(see discussion below), a computationally cheaper density estimator can usually be used.

\subsection{Sources of inaccuracy}
\label{sec:inaccuracies}

The accuracy of this method can be affected by a number of different factors.
Each of these potential sources of error will be tested independently in
Section \ref{sec:analtests}.

i) The gravitational force calculation in self-gravitating SPH codes does not simply
consist of summing
pairwise interactions.  In particular, the usual
gravitational force is typically  
``softened'' for small separations and distant particles are grouped together
in a `tree' structure, e.g. an oct-tree \citep{TreeGrav}. 
How aggressively particles are grouped and the extent of the gravitational 
softening is controlled by the user.  As our method relies
on the gravity estimated from the tree code, inaccuracies in the
gravitational force may propagate 
in our method also.

ii) The discrete realisation of a given density distribution may produce 
accelerations that differ from those produced by a continuous density field
since quantising the density
distribution inevitably introduces ``clumpiness'', which affects the resulting 
gravitational acceleration.  
Note that this error decreases with increasing resolution.  The gravitational
softening mentioned above also mitigates this effect.

iii) No physical disc is really an infinite slab of constant height, with 
density structure depending only on $z$.  Deviations from this structure
will obviously influence the accuracy of our method.

iv) Any inaccuracy in our map of surface densities $\Sigma_{map}$ will lead to
inaccuracies in our final estimate of $\Sigma$.  This inaccuracy will arise
from a combination of interpolation error (as the map is only calculated for
10\% of particles) and error due to the calculation of the surface density
itself.

To quantify the size of these relative errors, we note that effects i)-iii)
produce errors in the estimate of $\Sigma'$, while iv) produces errors in
$\Sigma_{map}$.  We represent the size of these errors as $\delta \Sigma'$ and
$\delta \Sigma_{map}$, respectively.  We then see that the relative error in
our estimate of $\Sigma$ is:

\begin{equation}
  \delta \left( \frac{\Sigma_{estim}}{\Sigma} \right) = \pm
  \frac{\sqrt{\delta \Sigma'^2 + \delta\Sigma_{map}^2}}{|\Sigma|}
  \label{eq:fullerror}
\end{equation}

Since $\Sigma$ is \emph{always} smallest at large $z$, it follows that the
fractional inaccuracy of our estimate is generally worst far from the
mid-plane.

\section{Analytic tests}
\label{sec:analtests}

As outlined
in the previous section, there are a number of possible sources of
inaccuracy in our column density estimate.  These are:
inaccuracies in the gravitational tree code, the quantization of a
continuous system, the ``infinite slab'' approximation and inaccuracy of the
surface density map.

In this section we construct a series of discs for which the gravitational
acceleration, density and
column density can be calculated analytically or semi-analytically. As described 
in Appendix \ref{sec:analiticCol},
these discs have power law density profiles between adjustable inner and
outer radii, with a Gaussian variation of density with $z$ at each radius.
The scale height of this Gaussian is adjusted so that it corresponds to
the hydrostatic equilibrium profile of a locally isothermal
non-self gravitating disc where the temperature is a power law function
of radius.  Because we are given a continuous, analytic density distribution, we can
integrate it semi-analytically to calculate the resulting vertical
component of the gravitational acceleration (equation \ref{eq:gzanalytic}). Furthermore, 
we can analytically determine the gravitational acceleration for the continuous density
distribution under the infinite slab approximation (equation \ref{gravity} without
the term due to the star, where $\Sigma'$ is given by equation \ref{analcol}).
We also calculate the vertical component of
gravitational acceleration using the oct-tree on the discretized
distribution. We will henceforth refer to the
gravitational acceleration calculated by integration of the
continuous density distribution as the ``continuous 
gravitational acceleration'' and the value calculated using the oct-tree as the 
``oct-tree gravitational acceleration''. Combining these three estimates of $g_z$, 
we can independently test the sources of error identified in
section \ref{sec:inaccuracies}.  We also test the accuracy of our
total surface density map by comparing it to the analytic value given by
\ref{analcol}. 

We emphasise that these discs do not 
represent realistic physical conditions that would result from the steady
state evolution of accretion discs, but are designed to test our method for 
calculating the
column density.  As such, many of these discs will have non-uniform values of
the Toomre Q parameter \citep{ToomreQ}, significantly different from unity and 
would change
significantly if allowed to evolve with time.

The properties of our analytic discs are summarised in Table \ref{tab:discs}.
In each case the density structure (Equation \ref{rho}) corresponds to a
hydrostatic structure of uniform temperature with mid-plane density scaling as
$R^{-0.5}$.   In running the simulations, typical parameters for the tree 
gravity were used (adaptive
gravitational softening on the SPH smoothing length $h$, Barnes \& Hut opening angle
threshold of 0.3).  

\begin{table*}
  \centering
  \begin{tabular}{|l|c|c|c|c|}
	 Parameter & Thin Disc & Thick Disc & Small Disc & High resolution\\
    $H/R$  & $0.025\sqrt{R/R_i}$ & $0.76\sqrt{R/R_i}$ & $0.025\sqrt{R/R_i}$ & $0.025\sqrt{R/R_i}$\\
    $\frac{R_o-R_i}{R_o}$ & 0.95& 0.95&0.33&0.95\\
    $N$ & $10^5$ & $10^5$ & $10^5$ & $5*10^5$
  \end{tabular}
  \caption{The relevant properties for the gravitational force and column
  density, for different analytic disc geometries. $H/R$ is the aspect ratio 
  of the disc, given as a function of cylindrical radius, $R_o$ and $R_i$ are the
  disc's outer and inner radii and $N$ is the number of particles in the
  simulation.  All discs are run with $q = M_*/M_{disc} = 0.1$.}
  \label{tab:discs}
\end{table*}

\subsection{Varying scale heights}

We first test how the accuracy of our method depends on disc thickness,
where thickness is measured by $H/R$ and $H$ is the scale height of the disc.  
We consider a thin disc with $0.025 < H/R < 0.11$ and a thick disc with $0.76 < H/R < 3.4$.

\begin{figure}
  \begin{center}
	 \includegraphics{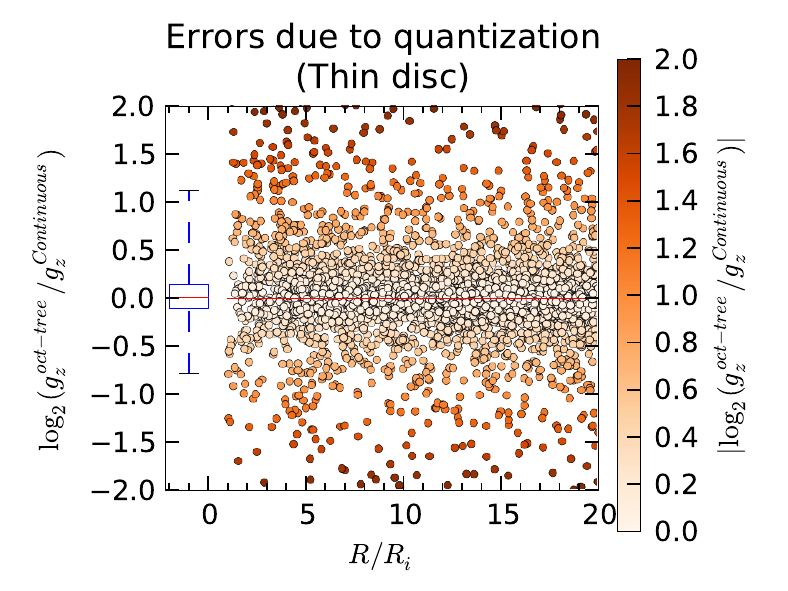}
  \end{center}
  \caption{The log2 ratio of the vertical component of the oct-tree and
  continuous 
  gravitational accelerations, as a 
function of cylindrical radius 
for a subset of 5000 particles in the `thin disc' calculation ($H/R=0.025\sqrt{R/R_i}$). 
 The boxplot on the left of the plot shows the distribution of errors.  The
 box bounds the central 50\% of particles (the edges are the 25\% and
  75\% percentiles), the red line is the median and the whiskers mark
  the 5\% and 95\% percentiles.}
  \label{fig:LowHQuant}
\end{figure}

\begin{figure}
  \begin{center}
	 \includegraphics{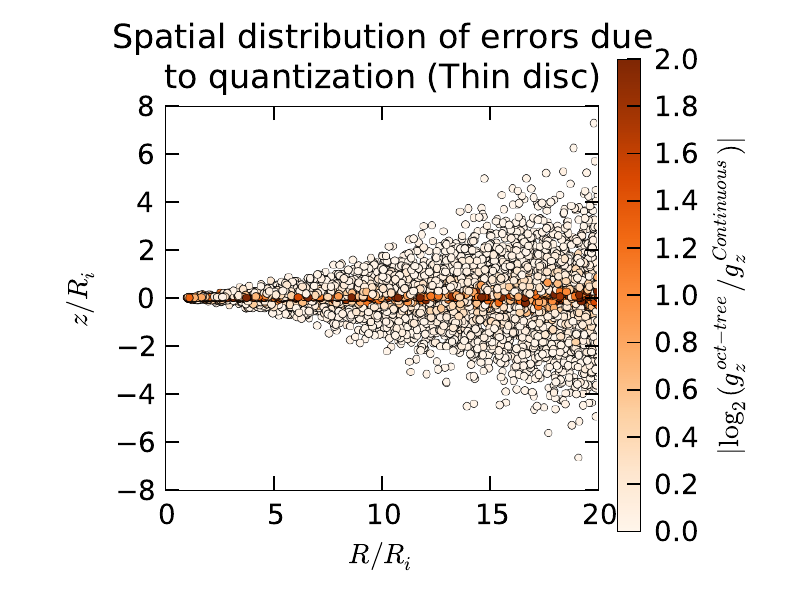}
  \end{center}
  \caption{The locations of the particles with errors in vertical component
of the gravitational acceleration colour coded as in Figure \ref{fig:LowHQuant}.}
  \label{fig:LowHQuantDist}
\end{figure}

\begin{figure}
  \begin{center}
	 \includegraphics{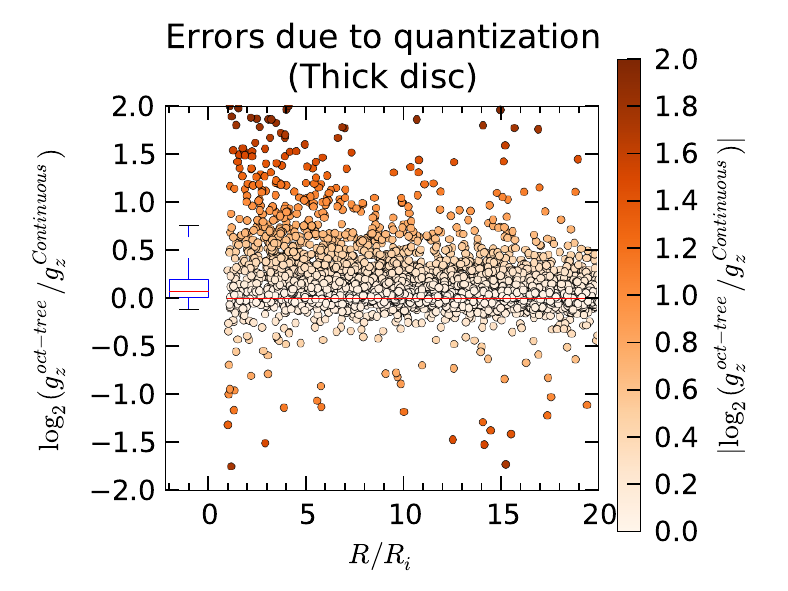}
  \end{center}
\caption{The log2 ratio of the vertical component of the oct-tree and
continuous gravitational accelerations, as a function of cylindrical radius 
for a subset of 5000 particles in the `thick disc' 
calculation ($H/R=0.76\sqrt{R/R_i}$).
The boxplot represents the distribution of errors on this plot (see Figure \ref{fig:LowHQuant} 
for explanation).}
  \label{fig:LargeHQuant}
\end{figure}

\begin{figure}
  \begin{center}
	 \includegraphics{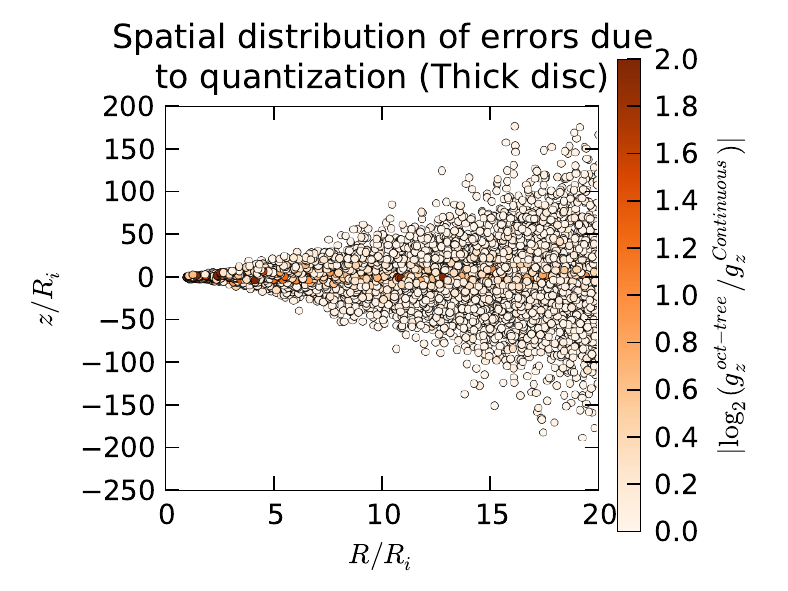}
  \end{center}
 \caption{The locations of the particles with errors in vertical component
of the gravitational acceleration colour coded as in Figure \ref{fig:LargeHQuant}.}
  \label{fig:LargeHQuantDist}
\end{figure}

Figures \ref{fig:LowHQuant} and \ref{fig:LargeHQuant} show 
the inaccuracy in the vertical component of gravitational acceleration
due to making the continuous density distribution (equation \ref{rho})
discrete and demonstrate that, as expected, 
the errors are highest in thinner discs, particularly close
to the disc mid-plane where any given particle is most affected by the 
``lumpy'' nature of its  environment (see Figures \ref{fig:LowHQuantDist} and
\ref{fig:LargeHQuantDist}).  The effect is exacerbated by the fact that
$g_z\rightarrow 0$ in the densest parts of the disc, so the relative errors
are largest here.  Note that such quantization errors are reduced by 
increasing the resolution  (see section \ref{sec:resTest}).

\begin{figure}
  \begin{center}
	 \includegraphics{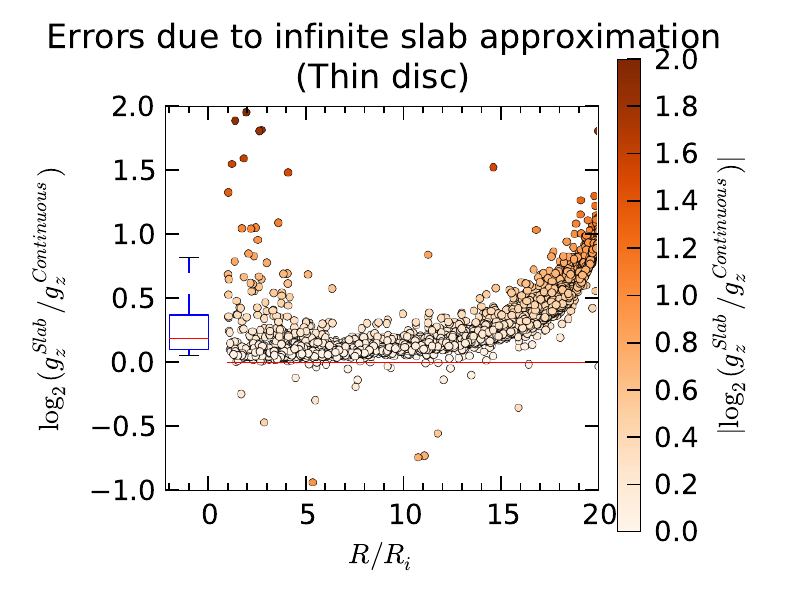}
  \end{center}
  \caption
  {The log2 ratio of the vertical component of the
gravitational acceleration calculated using the infinite slab approximation 
to the ``continuous'' value of the same quantity is plotted as a function of 
cylindrical radius  
  for a subset
of 5000 particles 
in the `thin disc' calculation ($H/R=0.025\sqrt{R/R_i}$).  
The boxplot represents the distribution of errors on this plot (see Figure \ref{fig:LowHQuant} 
for explanation).}
  \label{fig:LowHSlab}
\end{figure}

\begin{figure}
  \begin{center}
	 \includegraphics{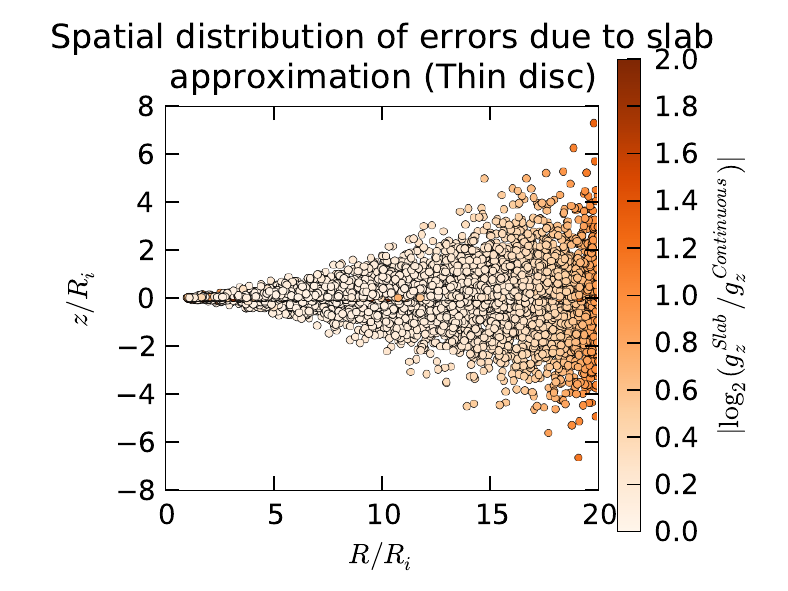}
  \end{center}
 \caption{The locations of the particles with errors in the vertical component
of the gravitational acceleration due to the infinite slab approximation
colour coded as in Figure \ref{fig:LowHQuant}.}
  \label{fig:LowHSlabDist}
\end{figure}

\begin{figure}
  \begin{center}
	 \includegraphics{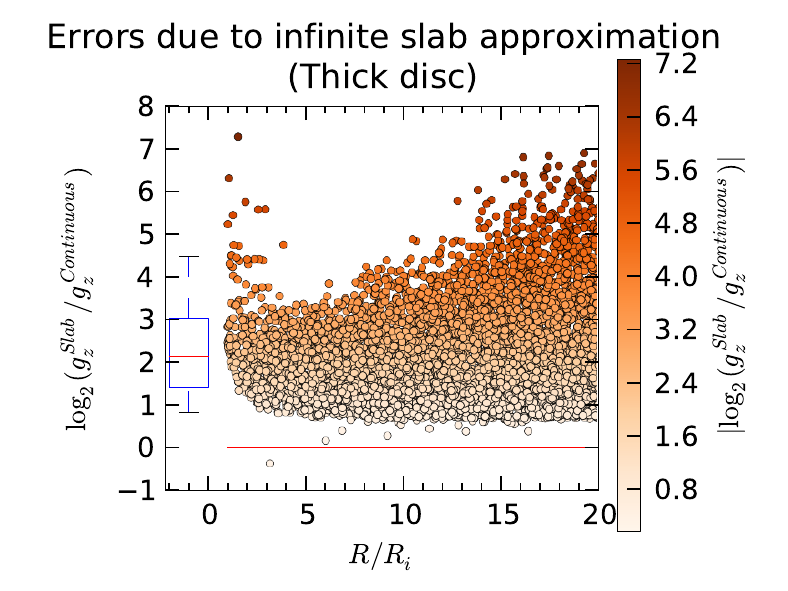}
  \end{center}
  \caption
  {The log2 ratio of the vertical component of the
gravitational acceleration calculated using the infinite slab approximation to
the ``continuous'' value of the same quantity
is plotted as a function of cylindrical radius for a subset
of 5000 particles in the `thick  disc' calculation ($H/R=0.76\sqrt{R/R_i}$).
The boxplot represents the distribution of errors on this plot (see Figure \ref{fig:LowHQuant} 
for explanation).}
  \label{fig:LargeHSlab}
\end{figure}

\begin{figure}
  \begin{center}
	 \includegraphics{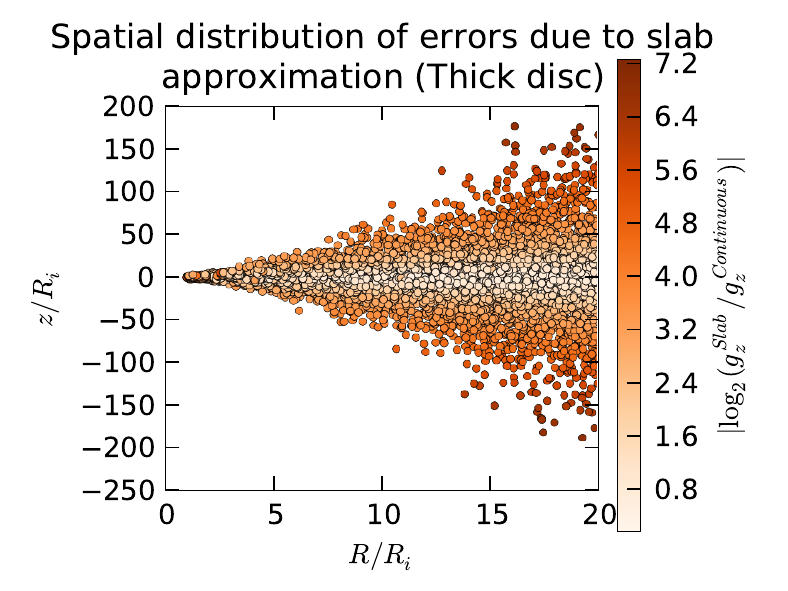}
  \end{center}
  \caption{The locations of the particles with errors in vertical component
of the gravitational acceleration colour coded as in Figure \ref{fig:LargeHSlab}}
  \label{fig:LargeHSlabDist}
\end{figure}

Next we investigate the inaccuracy in the gravitational acceleration due to 
use of the infinite slab approximation.  Figure \ref{fig:LowHSlab} show that this error 
is small for the thin disc, with most points being accurate to within 50\% of 
the expected analytic value.  Figure \ref{fig:LargeHSlab} shows that the accuracy of this
approximation breaks down significantly in the thick disc limit.  Note that
this disc has been chosen to test the thick disc limit and does not represent a
realistic physical situation.  Figures \ref{fig:LowHSlabDist}
 and \ref{fig:LargeHSlabDist} show a strong trend towards
greater inaccuracy where the disc is thickest, since it is at large $z$
that particles `see' the gravitational influence of radial
gradients in the disc.  The accuracy also decreases at the disc's edges where
the finite nature of disc has the largest effect.

\begin{figure}
  \begin{center}
	 \includegraphics{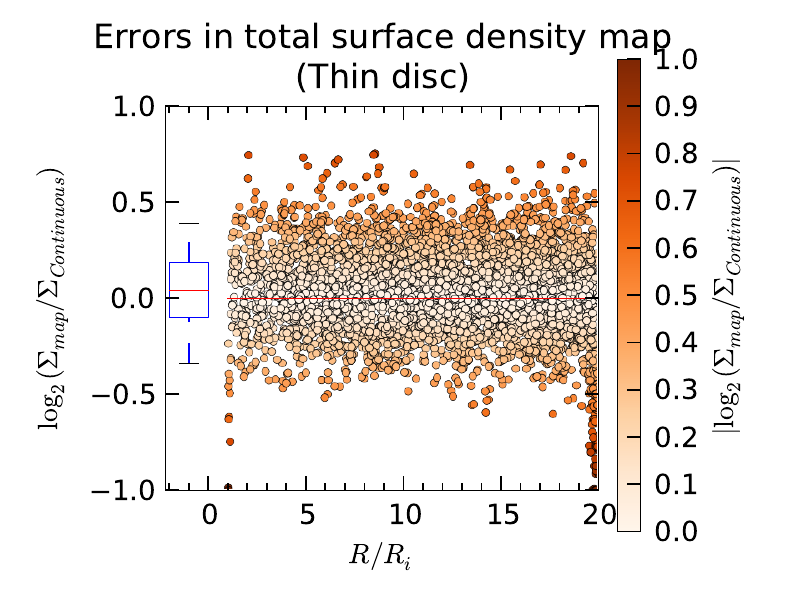}
  \end{center}
  \caption
 { The log2 ratio of the calculated total surface density map to the known
  analytic values is shown as a function of radius 
for 5000 particles from the `thin disc' simulation. 
The boxplot represents the distribution of errors on this plot (see Figure \ref{fig:LowHQuant} 
for explanation).}

  \label{fig:LowHSur}
\end{figure}

\begin{figure}
  \begin{center}
	 \includegraphics{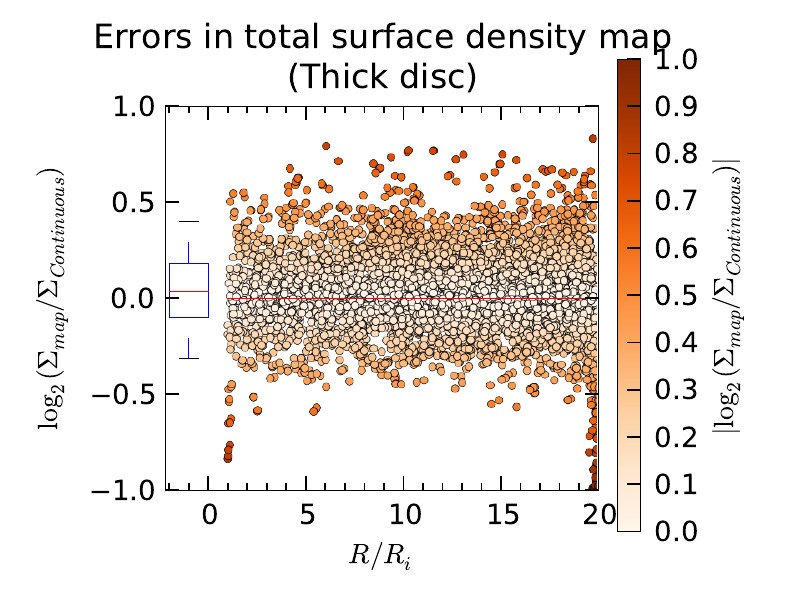}
  \end{center}
  \caption{The log2 ratio of the calculated total surface density map to the known
  analytic values is shown as a function of radius
for 5000 particles from the `thick disc' simulation. See Figure
The boxplot represents the distribution of errors on this plot (see Figure \ref{fig:LowHQuant} 
for explanation).}

  \label{fig:LargeHSur}
\end{figure}

Figures \ref{fig:LowHSur}
and \ref{fig:LargeHSur} test the accuracy of the total surface density map for the thin 
and thick discs, i.e. we here compare the result of interpolating the column
density map obtained by sampled measurements of the projected particle
distribution with the analytic value.  It is clear that the accuracy is
relatively insensitive to disc thickness and that most points are
accurate to within a few 10s of \%, with a median accuracy of
$\sim 10\%$ for both discs.

Overall, we find that the accuracy of our estimation of the column density to the surface
is controlled by errors in determining the column density to the mid-plane, rather than
by errors in the total surface density. The main source of the error in
determining the column density to the mid-plane depends on disc thickness (and
resolution, see \ref{sec:resTest}): thin discs are dominated by discreteness effects 
whereas in thick discs the accuracy of the infinite slab approximation is the limiting
factor.

\subsection{Effect of resolution}
\label{sec:resTest}

On theoretical grounds, we expect that the error due to quantisation will
roughly depend on the ``extra'' acceleration imparted on a particle by its
neighbours.  Given that all particles are identical, the mass of each particle
is given by $m = M_{disc}/N$, where $N$ is the number of particles in the
simulation.  Furthermore, the mean particle separation, $\lambda$, will scale
roughly as $\lambda \propto N^{-1/3}$ in a three-dimensional environment.
Therefore, we expect the inaccuracy due to quantization to scale as:

\begin{equation}
  \frac{Gm}{\lambda^2} \propto \frac{1}{N^{1/3}}
  \label{eq:quantScaling}
\end{equation}

This effect will be mitigated by the inclusion of gravitational softening,
which for a typical SPH simulation will modify gravity so the effect of the
quantized particles becomes:

\begin{equation}
  \frac{Gm}{(\lambda+\varepsilon)^2}
  \label{eq:quantSoftening}
\end{equation}

\noindent where $\varepsilon$ is typically set to the SPH smoothing length $h$ and $h$ is given by,

\begin{equation}
  \rho h^3 = \nu^3 m
  \label{eq:h}
\end{equation}

\noindent where $\nu$ is a numerical parameter usually set to 1.2.  Combining
equations \ref{eq:quantSoftening} and \ref{eq:h} we see that the inclusion of 
gravitational softening does not change
the $N^{-1/3}$ dependence of the quantization error, it merely decreases it by
a factor of $1/(1+\nu)^2 \sim 0.25$, compared to no gravitational softening.

\begin{figure}
  \begin{center}
	 \includegraphics{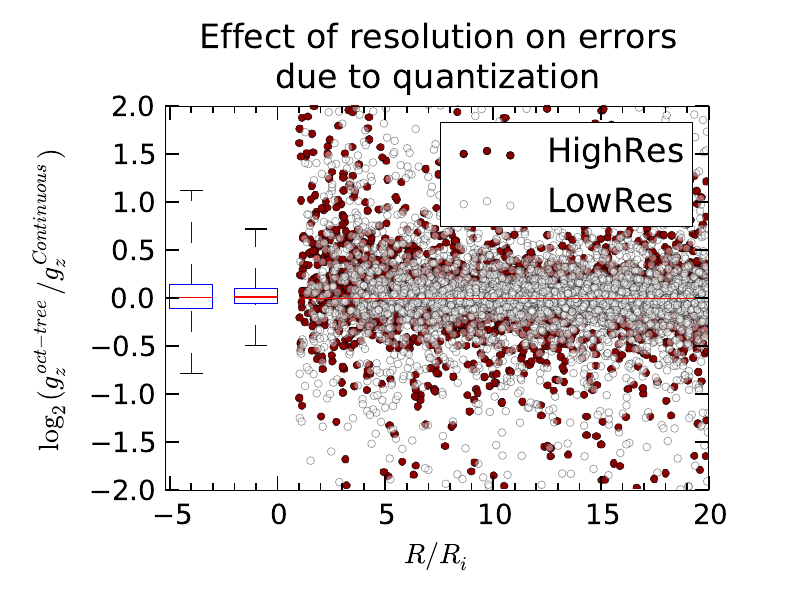}
  \end{center}
  \caption{The log2 ratio of the vertical component of the oct-tree to the 
  continuous gravitational acceleration for a subset of 
5000 particles as a function of cylindrical radius in the 'thin disc'
calculation ( $H/R=0.025\sqrt{R/R_i}$) at two different resolutions ($10^5$ and
$5*10^5$ particles).  The left and right boxplots show the low and high
resolution data respectively.  
The boxplots represents the distribution of errors on this plot (see Figure \ref{fig:LowHQuant} 
for explanation).}

    \label{fig:ResComp}
\end{figure}

In order to validate this predicted dependence on resolution, we repeat the thin
disc simulation with 5 times as many particles ($5*10^5$).  The two runs are 
compared in Figure \ref{fig:ResComp}, which shows that inaccuracy decreases 
as $N$ increases. We calculate the median relative error for both the high and
low resolution simulations and find they have a ratio of $\sim 1.6$.  Comparing 
this with equation \ref{eq:quantScaling}, which predicts a decrease in error 
$\approx 5^{1/3} = 1.7$, we see that our simulations are consistent with our
theoretical predictions. All other comparisons have not been shown as they are 
essentially unchanged with resolution.

We see that for both our $N=10^5$ and $N=5*10^5$ runs, the error in column
density estimate to the mid-plane is dominated for most particles by the
inaccuracy of the infinite slab approximation (compare boxplots in Figure
\ref{fig:ResComp} with that in Figure \ref{fig:LowHSlab}) rather than by
(resolution dependent) errors in computing $g_z$.  Considering the region at
$R/R_i \sim 16$ (where $H/R \sim 0.1$ in the thin disc, a value typical of
protostellar discs) we can estimate (using the above $N^{-1/3}$ scaling of
discreteness errors) that $N$ would have to be reduced to of order $10^4$
before discreteness errors dominated.  Such a low value of $N$ would never be
employed in a hydrodynamic calculation in any case, since it would imply that
the smoothing length was larger than the disc's vertical scale height (see
\cite{honHRes} and references therein).  We therefore conclude that our method
of estimating the column density to the mid-plane imposes no extra resolution
requirements on the code.

\subsection{Other parameters: disc mass and radial extent}

The accuracy of our method is independent of the ratio of the disc mass
to central object mass since the gravitational contribution due to the
star is not used by our method. Changing the radial extent of the disc means that
edge effects affect the accuracy of the infinite slab approximation - for example, 
we found that a radially restricted disc (with fractional width of $\sim 0.33$ compared
with $\sim 0.95$ in the simulations discussed above) increased such errors by a
factor of a few.

Generally, the infinite slab approximation is very accurate for
most particles in the simulation, with 90\% of particles being accurate to
within a factor of 2 and over 50\% of particles accurate to within a few 10s of percent.
Although the inaccuracy is acceptable, we find
that the infinite slab approximation slightly over-estimates the
gravitational acceleration in a systematic way.  Furthermore, the approximation
does worst at the edges of the disc (both radially and vertically), where the deviation from an 
infinite slab is most obvious.

The accuracy of the total surface density map produced by
our method is high, higher even than the accuracy of the infinite slab
approximation, which is already very good.  As such, we do not expect the
total surface density map to be a significant source of error in any application.

\subsection{Accuracy of the column density to the surface estimate.}

\begin{figure}
  \begin{center}
	 \includegraphics{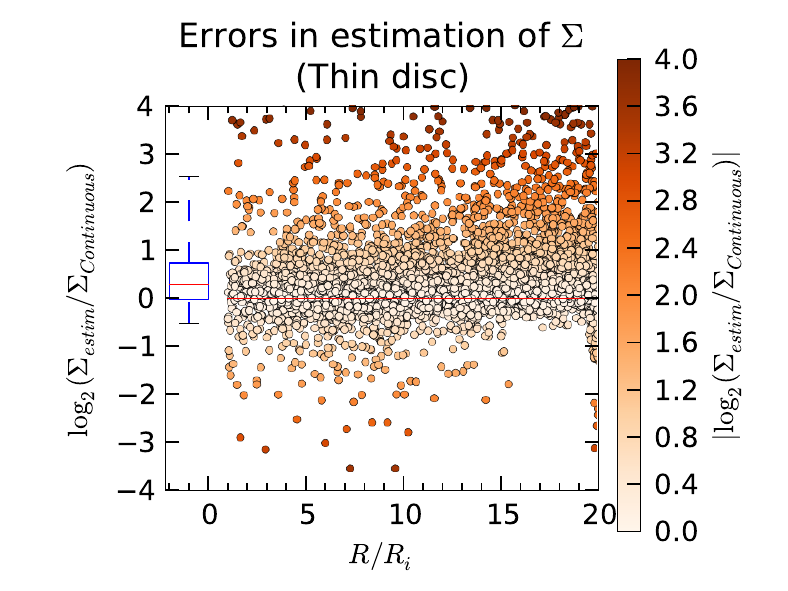}
  \end{center}
  \caption{The relative error in $\Sigma$ (the column density to the surface
  from a particle) is plotted for a subset of 5000 particles as a function of
  cylindrical radius in the 'thin disc' calculation ($H/R=0.025\sqrt{R/R_i}$).
The boxplot represents the distribution of errors on this plot (see Figure \ref{fig:LowHQuant} 
for explanation).}

  \label{fig:LowHCDsur}
\end{figure}

\begin{figure}
  \begin{center}
	 \includegraphics{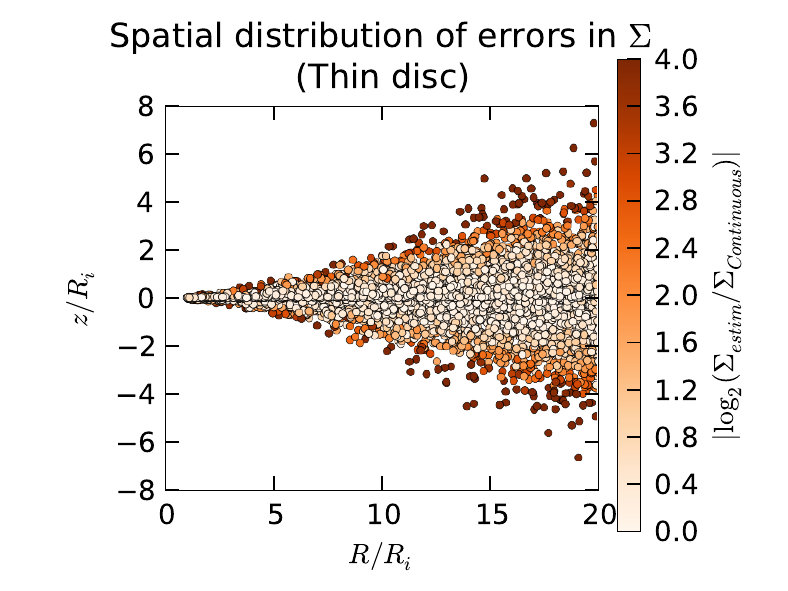}
  \end{center}
  \caption{The locations of the particles with errors in $\Sigma$ (the column
  density to the surface from a particle) colour coded as in Figure
  \ref{fig:LowHCDsur}.}
  \label{fig:LowHCDsurDist}
\end{figure}

Having dissected the various sources in error in calculating total surface
densities and column densities to the mid-plane, we can understand the
over-all accuracy of our determination of $\Sigma$ (the column
density from each particle to the surface) which is to be used in the 
calculation of optical depths and cooling rates. 
Figures \ref{fig:LowHCDsur} and \ref{fig:LowHCDsurDist} show the relative 
error in $\Sigma$ (the column density
from the particle to the surface), and the spatial distribution of these
errors.  

It is evident that the over-all accuracy is good (most particles are accurate
to within a factor 2) and that the largest errors are found at large $z$. This is 
to be expected since for particles
close to the mid-plane, the column density to the surface is controlled by the
total surface density whose accuracy is of order $10 \%$,  regardless of
the accuracy of determination of the column density to the mid-plane.
At high $z$, the column density to the surface is obtained by the subtraction
of two quantities of similar magnitude and the result is particularly sensitive
to errors in determining the column density to the mid-plane (which, at large
$z$, derive from the breakdown of the infinite slab approximation). 

In practice, however, inaccuracies at high $z$ may be irrelevant
 to the computation of the cooling rate according to equation \ref{eq:cooling}. This is
because once the disc enters the optically thin regime, the cooling rate
according to equation \ref{eq:cooling} is in any case independent of optical depth.
{\footnote{This is of course not to say that equation \ref{eq:cooling} is necessarily
a good approximation to cooling in the optically thin surface layers of
the disc: indeed flux limited diffusion is also a poor description of the
local cooling in such a layer and neither equation \ref{eq:cooling} nor flux-limited
diffusion should be used in applications where the cooling of surface layers
is of particular interest.}}

\section{Application to realistic discs}
\label{sec:realdiscs}

Up until now we have focused on evaluating our approximation in discs for which the 
column density could be analytically determined, rather than those that represent the 
outcome of realistic disc evolution.  If our method is to be of any use, it is important 
that it is accurate not only in idealised problems, but in ``real world'' simulations.  
To provide such a test, we evaluate our method on a disc simulation kindly
provided by Ken Rice and detailed in Section 3 of \citep{DataPaper} (Simulation 1, Table 1).
Briefly, the disc was evolved using SPH with radiative cooling implemented
using the method describe in \citep{HybridCooling}.  The disc was constructed
using $5*10^5$ particles distributed between 10 AU and 50 AU with a
surface density profile $\Sigma \propto R^{-3/2}$.  The disc to star mass ratio, $q$,
was set to $0.25$.  The simulation was run for 27 outer rotation periods\footnote{An outer
rotation period is defined as the rotation period at the initial outer radius
of the disc, which here is 50 AU, leading to an outer rotation period of 354
yrs.}, which was long enough for the disc to develop spiral structures and settle into marginal
stability ($Q \sim 1$), where radiative cooling was matched by heat generated through
gravitational instabilities.

Unlike the analytic discs, we cannot evaluate the effects of quantization and
the infinite slab approximation separately.  Therefore, we test our method by
comparing our estimated column densities to the mid-plane and the surface, to 
those obtained using a counting method described in Appendix \ref{sec:counting}.

Figure \ref{fig:SpiralMid} shows a comparison of the gravity based estimates of
the column density to the mid-plane and the results of the counting method.
Even though this comparison includes inaccuracies from both the
infinite slab approximation and the quantization of the disc, the overall
accuracy remains extremely high.  In fact, over 90\% of particles agree to
within a factor of two, with most particles agreeing to within a few 10s of
percent.  Furthermore, Figure \ref{fig:SpiralMidDist} shows that those particles with high
inaccuracy are located in surface layers where they are unlikely to 
affect calculations of the cooling rate.

\begin{figure}
  \begin{center}
	 \includegraphics{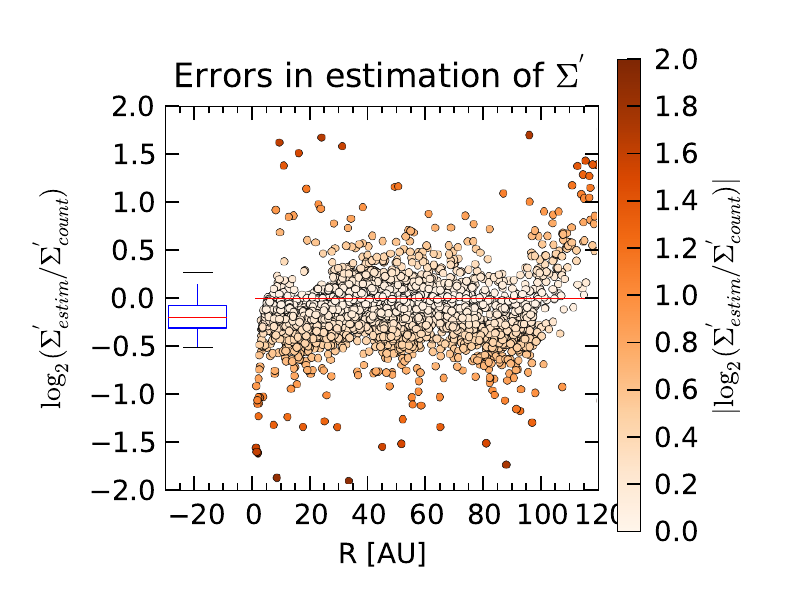}
  \end{center}
  \caption{The log2 ratio of $\Sigma'$ (the column density between a particle
  and the \emph{mid-plane}) calculated using our method and by
  direct counting for a subset of 5000 particles as a function of cylindrical
  radius for the `marginally stable' disc (Simulation 1, \citep{DataPaper}).
The boxplot represents the distribution of errors on this plot (see Figure \ref{fig:LowHQuant} 
for explanation).}

  \label{fig:SpiralMid}
\end{figure}

\begin{figure}
  \begin{center}
	 \includegraphics{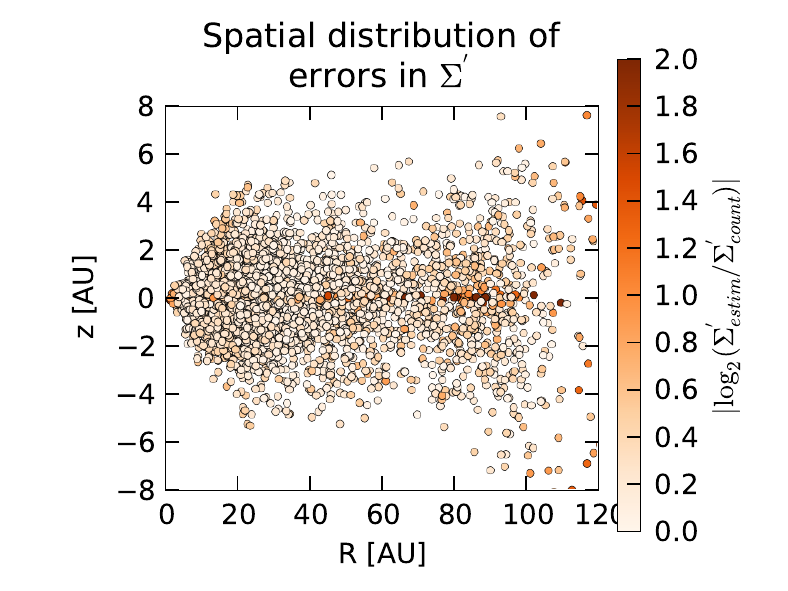}
  \end{center}
  \caption{The locations of the particles with errors in $\Sigma'$ colour coded
  as in Figure \ref{fig:SpiralMid}.}
  \label{fig:SpiralMidDist}
\end{figure}

\begin{figure}
  \begin{center}
	 \includegraphics{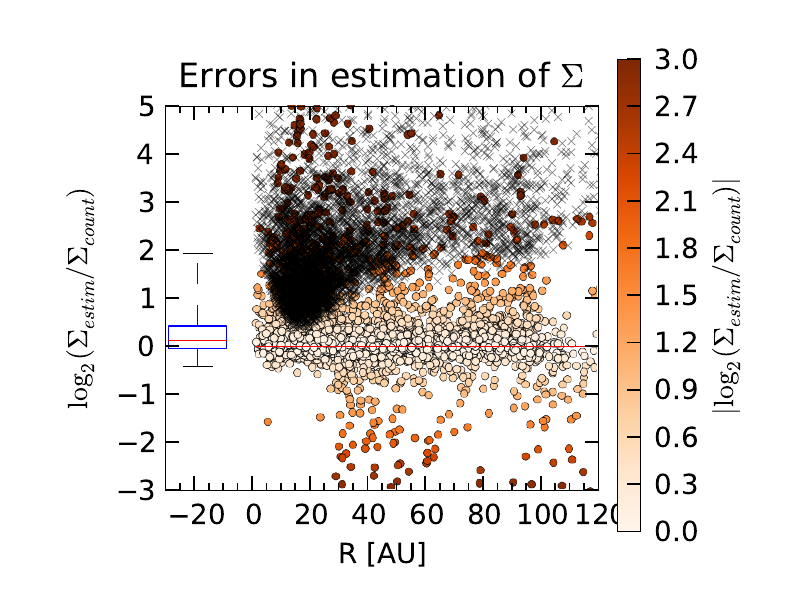}
  \end{center}
  \caption{The log2 ratio of $\Sigma$ (the column density between a particle and the surface)
  calculated using our method and by direct
  counting, for a subset of 5000 particles as a function of cylindrical radius
  for the `marginally stable' disc (Simulation 1, \citep{DataPaper}).  The black
  crosses show the log2 ratio of estimating $\Sigma$ using the Stamatellos
  et al method \citep{PolytropicCooling} and the same value calculated by
  direct counting. 
The boxplot represents the distribution of errors on this plot (see Figure \ref{fig:LowHQuant} 
for explanation).}

  \label{fig:SpiralSur}
\end{figure}

\begin{figure}
  \begin{center}
	 \includegraphics{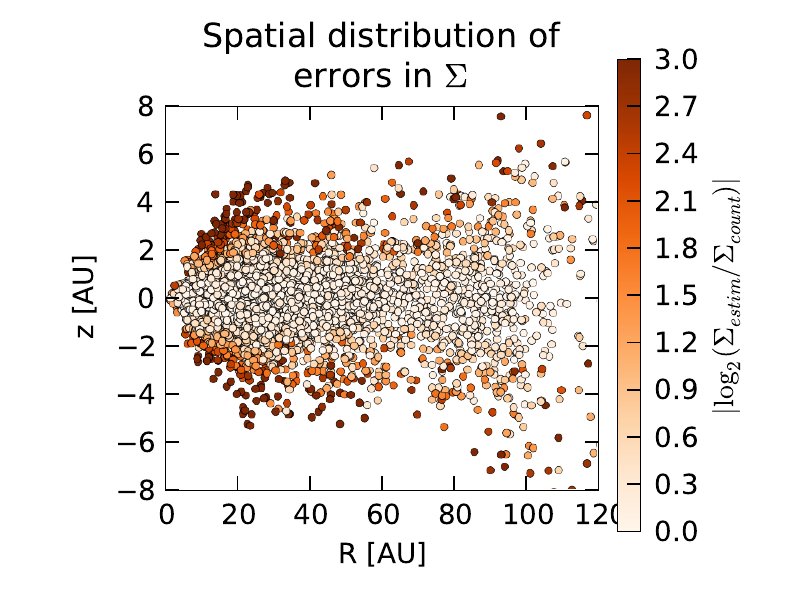}
  \end{center}
  \caption{The locations of the particles with errors in $\Sigma$ colour coded
  as in Figure \ref{fig:SpiralSur}.}
  \label{fig:SpiralSurDist}
\end{figure}

Figure \ref{fig:SpiralSur} shows the accuracy of the column density to the surface,
$\Sigma$, obtained by subtracting $\Sigma'$ from $\Sigma_{map}$.  As expected, 
the use of our
total surface density map to infer column density to the surface does not significantly 
increase the error.  The spatial distribution of the errors, shown in Figure
\ref{fig:SpiralSurDist}, show that the inaccuracy in our method only becomes 
significant in the optically thin region, where it is unimportant for our 
calculation of the cooling rate. 

For comparison, the column densities to the surface 
estimated by the Stamatellos et al \citep{PolytropicCooling} method are also shown 
in black in Figure \ref{fig:SpiralSur}\footnote{Only the disc's potential was used in calculating the Stamatellos estimate.}. The discrepancy is less large than that
found by \cite{WilkinsAndClarke} (who used a thinner disc), but is significant
nonetheless.  Specifically, the Stamatellos method has a larger dispersion than
our method and has a systematic offset of around a factor of 4.  As can be seen
in equation \ref{eq:limitCase}, the cooling rate depends on $\Sigma$
quadratically in the optically thick limit, so this factor 4 discrepancy gives
an order of magnitude under-estimate of the cooling rate.

\begin{figure}
  \begin{center}
	 \includegraphics{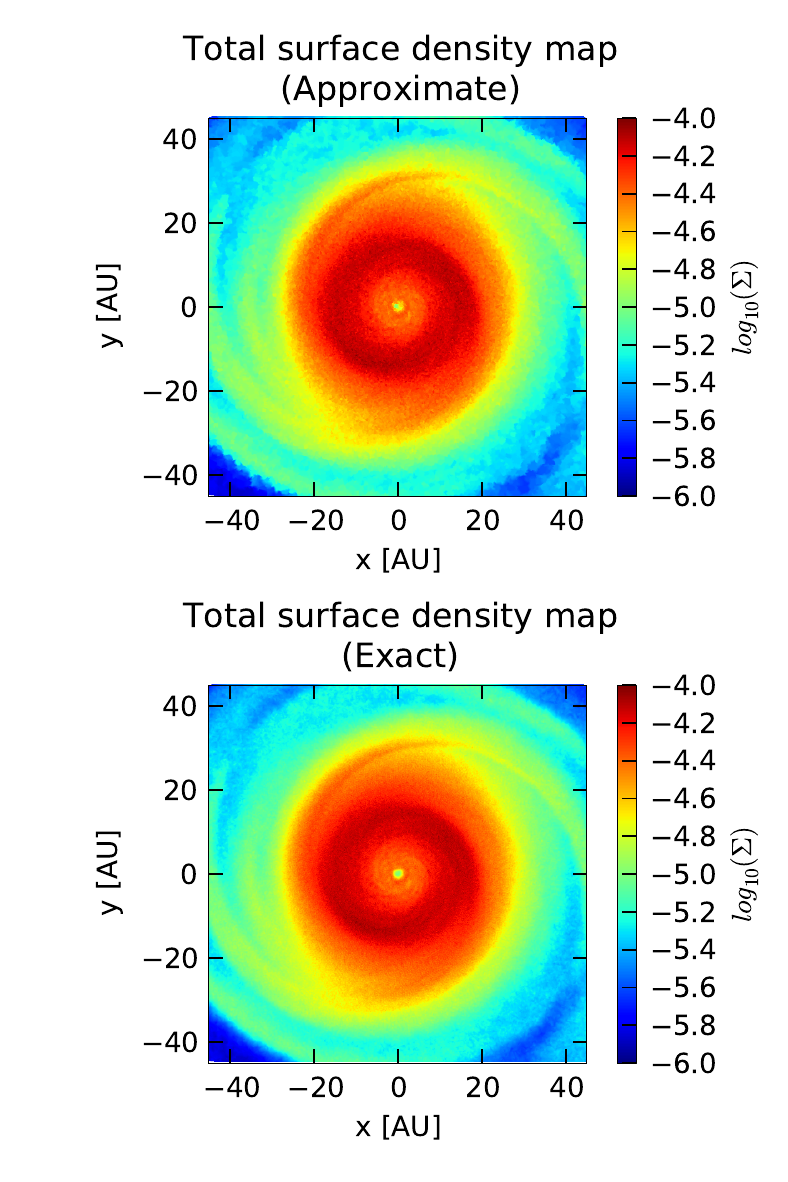}
  \end{center}
  \caption{The two heat maps show the total surface density inferred by the method
  used to estimate the column density (top) and an ``exact'' counting method
  (bottom).  The colour bars show the mapping between total surface density and
  colour used in this plot (log scale).}
  \label{fig:denmap}
\end{figure}

For simulations involving spiral structures, such as the one considered here, it 
is important to be able to resolve the differences between the cooling rates in 
the spiral arms and outside of them.  In Figure \ref{fig:denmap} we plot our
estimated total surface density map next to an ``exact map'' and show that the
essential features of the disc are still clearly visible.  This implies that our 
method would be able to 
correctly distinguish between on arm and off arm cooling.  We find that
$\delta\Sigma/\Sigma < 0.05$ for $R < 20$ and then rises steadily to a
mean value of $\sim 0.2$.  Comparing this with Figure \ref{fig:denmap}, we
conclude that we can easily resolve density perturbation for which
$\delta\Sigma/\Sigma > 0.05$.  Resolving smaller density perturbations could be
achieved by increasing from 10\% the number of points at which the surface
density map is calculated\footnote{Note the choice of 10\% of particles is
somewhat arbitrary.  Spiral structure can still be resolved even if the column
density is calculated at only 1\% of particles (data not shown).  This is
discussed in more detail in Appendix \ref{sec:surfaceDen}}.

\subsection{Fragmented discs}
\label{sec:fragdisc}

\begin{figure}
  \begin{center}
	 \includegraphics{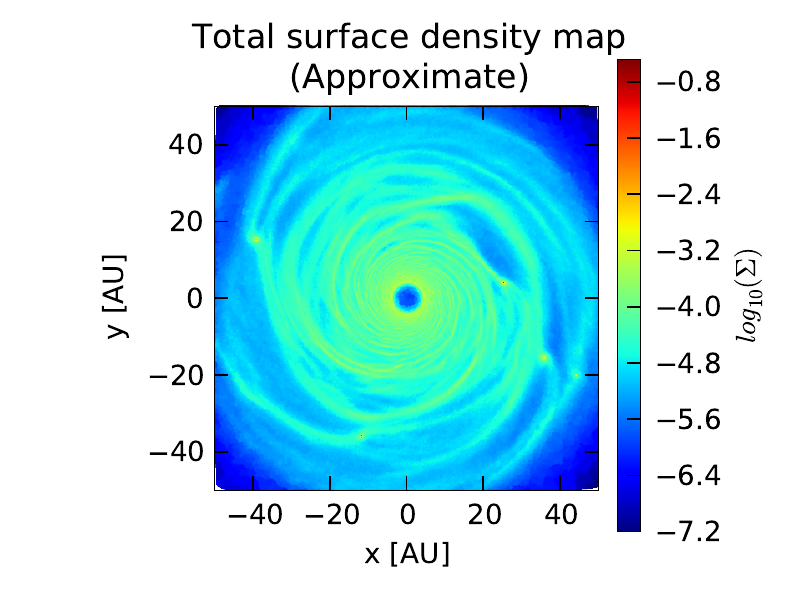}
  \end{center}
  \caption{A heat map showing the total surface density estimate for the fragmented 
  disc described in section \ref{sec:fragdisc}.  The colour bar shows the
  mapping between total surface density and colour on a log scale.}
  \label{fig:fragsurface}
\end{figure}

While the above section shows that our method works
well for self-gravitating discs in marginal stability (where heating due to gravitational
turbulence is balanced by cooling), it does not test its
ability to recover the cooling of fragments if they form.  To test this, we
consider a fragmented disc, which is originally 50 AU  in size, with a disc to star mass 
ratio of .1, which has been evolved for ~5 outer rotational periods (data provided
by Peter Cossins).  Figure \ref{fig:fragsurface} shows a total surface density map calculated using
our method, which clearly shows the presence of several fragments.

\begin{figure}
  \begin{center}
	 \includegraphics{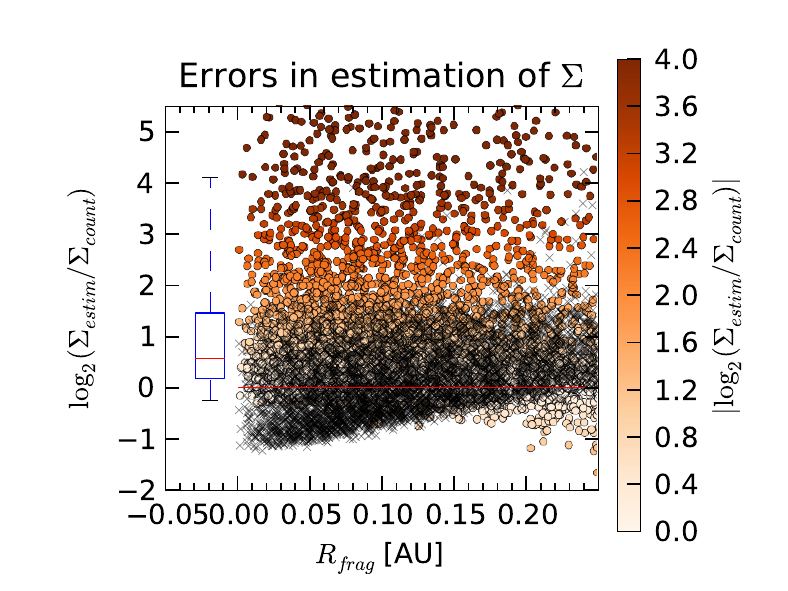}
  \end{center}
  \caption{The log2 ratio of $\Sigma$ (the column density between a particle and the surface)
  calculated using our method and by direct
  counting, for all particles within .25 AU of the fragment located at
  (-40,15) in Figure \ref{fig:fragsurface}, as a function of cylindrical radius from
  the centre of the fragment.  The black
  crosses show the log2 ratio of estimating $\Sigma$ using the Stamatellos
  et al method \citep{PolytropicCooling} and the same value calculated by
  direct counting. The boxplot represents the distribution of errors on this 
  plot (see Figure \ref{fig:LowHQuant} for explanation).}
  \label{fig:FragSur}
\end{figure}

\begin{figure}
  \begin{center}
	 \includegraphics{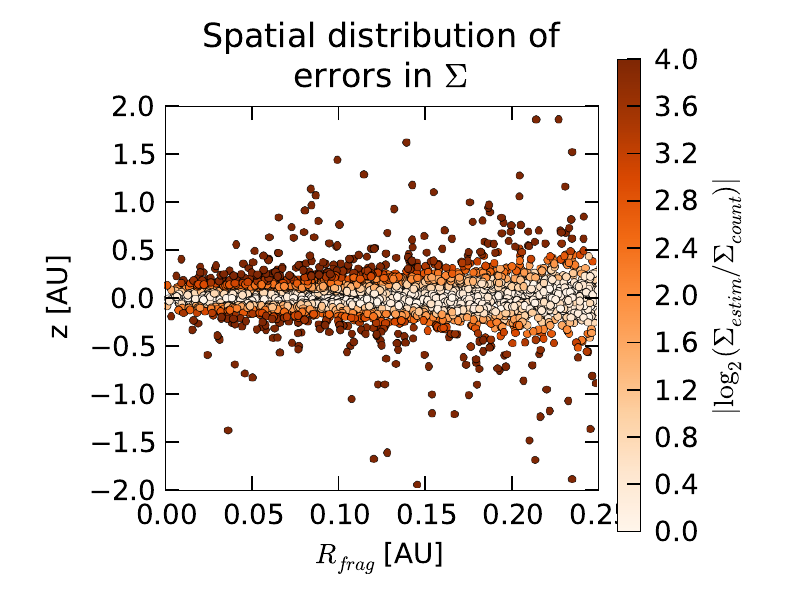}
  \end{center}
  \caption{The locations of the particles with errors in $\Sigma$ colour coded
  as in Figure \ref{fig:FragSur} as a function of cylindrical radius
  $R_{frag}$ and $z$ from the centre of the fragment.}
  \label{fig:FragSurDist}
\end{figure}

To test our methods ability to recover the column density within a fragment,
we focus the rest of our analysis on the .25 AU around the fragment at ~(-40,15) (see 
Figure \ref{fig:fragsurface}).  We calculate the column density to the surface
($\Sigma$) for each particle in this region and compare it to the column
density obtained using the same counting method as used in section
\ref{sec:realdiscs} above.  The results of
this comparison are shown in Figure \ref{fig:FragSur}.  As with Figure
\ref{fig:SpiralSur}, we also include the column density estimated by the
Stamatellos method for comparison (the black dots in the figure).  Figure
\ref{fig:FragSur} shows that our
method does not perform as well as the Stamatellos method in reproducing the
column density within fragments.  Although our method is still accurate to
within a factor of two for the majority of the particles, there is a
systematic trend for our method to over-estimate the column density, while the
Stamatellos method shows no systematically bias here, unlike the rest of the disc.  
Figure \ref{fig:FragSurDist} shows that as before, the lowest
accuracy points are still located at high $z$, where the column density is
less important to estimating the cooling rate.

It is perhaps unsurprising that our method performs less well within a
fragment, as fragments have a roughly spherical geometry where the infinite
slab approximation is likely to break down.  The same reasoning
explains why the Stamatellos method does so well here, as it is known to
perform well for problems with a spherical geometry.

The calculation of our total surface density map provides a
mechanism for combining the advantages of our method with those of the Stamatellos
method in fragmented discs.  That is, the Stamatellos method can be used to
estimate $\Sigma$ within the fragments and our method can be used everywhere else,
where a fragment can be defined to be a region where
the total surface density exceeds the average value by two orders of
magnitude.  Indeed, this is the exact definition used to identify fragments in
many studies of fragmenting discs (e.g. \cite{DataPaper}).  Such a scheme would
combine the advantages of the two methods and provide an estimate of the
column density without systematic bias for the entirety of the fragmented
disc and may be optimal when the cooling within fragments is an
important consideration.

\section{Conclusion}
\label{sec:conclusion}

In this paper we have presented a new method for efficiently and accurately
estimating the cooling in disc geometries.  To achieve this, our method
estimates the column density between each particle and the surface
of the disc, via estimates of the column density to the mid-plane and the
total surface density.  We verify the accuracy of our column density estimation (and
hence our cooling rates) by comparison with discs for which the analytic form
of the column density can be calculated.  We test our method on a
realistic proto-planetary disc simulation, that has been evolved for long enough to reach
marginal stability and develop a typical spiral structure.  Finally, we test
our method on a fragmented disc and find that our method does not perform as
well as the Stamatellos method within the fragments, due to the locally
spherical geometry.  We suggest that this shortcoming can be resolved by using
the Stamatellos method only within regions of extremely high density (i.e. the
fragments). We find throughout
our tests that the accuracy of our method remains high  (i.e. typical errors of
order a few tens of per cent) and conclude that it is ideally suited for
use in problems that depend on an accurate estimate of the cooling rate in disc
geometries.  

\section{Materials \& Methods}
\label{sec:materials}

In the interests of reproducibility and transparency, all code and data used
in performing this work have been made freely available online.

The generation of initial conditions and the code used to perform the analyses 
described in this paper can be found at 
\url{https://bitbucket.org/constantAmateur/disccolumndensity}.  See the
readme file in this repository for further details. The gravitational acceleration
for the analytic discs was calculated using a
modified version of GADGET-2.0 \citep{Gadget2Code}, which can be obtained from 
\url{https://bitbucket.org/constantAmateur/gadgetoutputgravaccel}.  Converting of
initial conditions to/from ascii files was done using the code from
\url{https://bitbucket.org/constantAmateur/easyic}.

The data from the spiral simulation used in section \ref{sec:realdiscs} were
provided by Ken Rice \citep{DataPaper}.  The fragmented disc used in section
\ref{sec:fragdisc} were provided by Peter Cossins and Giuseppe Lodato
(unpublished).  Both are made available at 
\url{https://bitbucket.org/constantAmateur/disccolumndensity} with the original 
authors' permission.

\section{Acknowledgements}
\label{sec:ack}

We would like to thank Ken Rice for providing the simulation data used is
section \ref{sec:realdiscs}.  We would also like to thank Giuseppe Lodato and
Peter Cossins for providing the simulation data use in section
\ref{sec:fragdisc}. Finally, we thank an anonymous referee for valuable
comments that have improved the paper.
Matthew Young gratefully acknowledges the support of a Poynton Cambridge Australia Scholarship.

\appendix

\section{The analytic column density}

\label{sec:analiticCol}
Consider a disc with density profile ($\rho (R,z)$) given by:

\begin{equation}
  \rho = \frac{M_d(\alpha_{\rho}+2)}{2\pi\left( R_o^{\alpha_{\rho}+2} -
  R_i^{\alpha_{\rho}+2}\right) } R^{\alpha_{\rho}}
  \frac{e^{-z^2/2H^2}}{\sqrt{2\pi H^2}}
  \label{rho}
\end{equation}

This corresponds to a disc of mass $M_d$, inner and outer radii $R_i$ and $R_o$
and surface density profile $\Sigma \propto R^{\alpha_\rho}$; the Gaussian
distribution with respect to $z$ corresponds to a situation of hydrostatic
equilibrium in the case that the disc is locally vertically isothermal, in
which case the scale height $H$ is given by:

\begin{eqnarray}
  H &=& \frac{c_s}{\Omega}\\
  c_s &=& \sqrt{\frac{\gamma k_b T}{\mu m_H}}\\
  \Omega &=& \sqrt{\frac{GM_*}{R^3}}
  \label{scaleH}
\end{eqnarray}

\noindent where $\gamma=5/3$ is the adiabatic index, $\mu = 2.3$ is the molecular weight,
$m_H$ is the mass of Hydrogen and $k_b$ is the Boltzmann constant.  Finally,
the temperature profile is given by

\begin{equation}
  T = T_0 \left(\frac{R}{R_i}\right)^{\alpha_T}
  \label{temp}
\end{equation}

\noindent Putting this all together gives:

\begin{equation}
  H = \sqrt{\frac{\gamma k_b T_0}{G M_* \mu m_H R_i^{\alpha_T} }}
  R^{\frac{\alpha_T +3}{2}}
  \label{eq:fullH}
\end{equation}

The vertical component of gravitational acceleration at $R$,$z$ is given by the following
integral:

\begin{equation}
  \int_{R_i}^{R_o}\int_{0}^{2\pi}\int_{-\infty}^{\infty}
  \frac{ G R' \rho (R',z') (z'-z) dR' d\theta' dz}{\left( R^2+R'^2 - 2 R R' \cos
  (\theta-\theta')+(z'-z)^2 \right)^{3/2}}
  \label{eq:gzanalytic}
\end{equation}

Since this integral cannot be solved analytically, we evaluated it
numerically using the ``NIntegrate'' function in mathematica using the
``Adaptive Monte Carlo'' method. The column density from each particle at
$R$,$z$ to the surface and to the mid-plane are obtained from integration of
equation \ref{rho}, i.e.

\begin{eqnarray}
  \Sigma' &=& \frac{M_d(\alpha_{\rho}+2)}{4\pi\left( R_o^{\alpha_{\rho}+2} -
  R_i^{\alpha_{\rho}+2}\right) } R^{\alpha_{\rho}} erf\left(
  \frac{z}{\sqrt{2H^2}} \right)\\
  \Sigma &=& \frac{M_d(\alpha_{\rho}+2)}{4\pi\left( R_o^{\alpha_{\rho}+2} -
  R_i^{\alpha_{\rho}+2}\right) } R^{\alpha_{\rho}} erfc\left(
  \frac{z}{\sqrt{2H^2}} \right)\\
  \Sigma_{total} &=& \frac{M_d(\alpha_{\rho}+2)}{4\pi\left(
  R_o^{\alpha_{\rho}+2} - R_i^{\alpha_{\rho}+2}\right) } R^{\alpha_{\rho}} 
  \label{analcol}
\end{eqnarray}

where $erf$ is the error function and $erfc$ is the complementary error
function.

\section{Calculating the total surface density map}
\label{sec:surfaceDen}

For our method to be useful, we have to be able to convert column densities to
the mid-plane to column densities to the surface without significant loss of
accuracy.  The most obvious way to do this is to try and use the fact that we
know the column density to the mid-plane for all particles and use the particles
at the top of the disc to approximate the surface density.  However, as is
shown in the main text, the accuracy of the column density to the mid-plane
estimates is lowest at the top of the disc.  As each estimate of the surface
density will potentially effect several particles below it, even small
inaccuracies will be compounded.  

In essence, what is required is a method for
estimating the two-dimensional surface density from the particle positions and
masses projected onto the mid-plane.  There are many such methods available,
each with advantages and disadvantages (see section \ref{sec:compeff} and
\cite{Ferdosi}).  Our method is not tied to any one technique for estimating
the total surface density and so the advantages and disadvantages of the
different techniques should be weighed against the scientific application of
interest.  In this paper, we adopt the following algorithm to calculate a
total surface density map.

\begin{enumerate}
  \item Randomly select 10\% of the particles, call these the tracer particles.
  \item Project all particles on the xy plane and for each tracer particle,
    find its 60 closest projected neighbours
  \item The column density at each tracer particle is then given by (mass of 60
    nearest particles)/$\pi R_{max}^2$ where $R_{max}$ is the distance between the
    tracer particle and its 60th furthest neighbour
\end{enumerate}

Because the tracer particles are chosen at random, the resolution of our surface
map automatically adjusts to the density profile of our disc.  Furthermore,
because we are calculating the surface density directly, the inaccuracies of
the gravity based estimates at large $z$ are not an issue\footnote{The accuracy
of the estimate of surface density at (x,y) could be improved by using the 2D
version of an SPH smoothing here.  However, this map will be used by many
points in the cylinder to represent the column density to the surface, so it is
preferable that our estimate be a little more ``washed out'' to increase the
accuracy for those particles off the (x,y) axis that also use this grid point as
an estimator of $\Sigma$.}.  This method has a computational complexity that
is roughly $O(NlogN)$.

An alternative of only slightly lower accuracy (data not shown), but significantly improved
computational efficiency $O(N)$, is to construct a two
dimensional grid and evaluate the column density by adding up the masses in
each grid cell and dividing by the grid area.  The grid can be spaced so that
roughly the same number of particles is located in each radial
annulus\footnote{This is done by calculating N equally spaced quantiles in the
distribution of cylindrical radii, a calculation which requires only a single
list sort operation.}, allowing the grid to adjust to the disc's density
profile.  The surface density for each particle is then given by the cell
within which it resides.

The exact computational cost of both the ``random sampling'' and ``grid''
methods suggested here will in general depend upon the code which is used for
the rest of the simulation (i.e. the hydrodynamics and gravity).  This is
because different codes have different costs associated with cross process
communication and different logical times where all particles in the simulation
are easily accessible.

\section{Calculating an ``exact'' column density}
\label{sec:counting}

For the realistic simulation considered in section \ref{sec:realdiscs}, the column 
density at each point to the surface or
the mid-plane cannot be determined analytically.  As such, we require some
method to calculate the column density that we can use as our gold standard
, to test the accuracy of our method.  To do this we use the same basic idea as 
employed in Appendix
\ref{sec:surfaceDen} to build the total surface density map.  In detail, we do the
following for each particle.
\begin{enumerate}
  \item Remove all particles from the simulation that are below the particle
    (for column density to surface) or not between the particle and the
    mid-plane (for column density to the mid-plane).
  \item Project all remaining particles onto the xy plane, find either the 60
    closest particles or the number of particles within a circle of radius 3
    AU,
    centred on the point for which we are trying to calculate the column
    density.
  \item The column density is then given by the sum of the masses of our
    neighbouring particles divided by $\pi R_{max}^2$ which $R_{max}$ is either
    the distance to the 60th particle (in the xy plane) or 3 AU if there are fewer
    than 60 neighbours within 3 AU.
\end{enumerate}

\bibliographystyle{mn2e}
\bibliography{references}
\end{document}